# Integrated Snapshot Near-infrared Hypersepctral Imaging Framework with Diffractive Optics

Jingyue Ma
State Key Laboratory of Information Photonics and Optical Communications
Beijing University of Posts and telecommunications
Beijing, China
ma_jingyue@bupt.edu.cn

Zhenming Yu*
State Key Laboratory of Information Photonics and Optical Communications
Beijing University of Posts and telecommunications and Xiong'an Aerospace Information Research Institute
Beijing, China
yuzhenming@bupt.edu.cn

Zhengyang Li
State Key Laboratory of Information Photonics and Optical Communications
Beijing University of Posts and telecommunications
Beijing, China
zhengyang.li@bupt.edu.cn

Linag Lin
State Key Laboratory of Information Photonics and Optical Communications
Beijing University of Posts and telecommunications
Beijing, China
aliang@bupt.due.cn

Liming Cheng
State Key Laboratory of Information Photonics and Optical Communications
Beijing University of Posts and telecommunications
Beijing, China
LMCheng@bupt.edu.cn

Kun Xu
State Key Laboratory of Information Photonics and Optical Communications
Beijing University of Posts and telecommunications and Xiong'an Aerospace Information Research Institute
Beijing, China
xukun@bupt.edu.cn

*Abstract*—We propose an integrated snapshot near-infrared hyperspectral imaging framework that combines designed DOE with NIRSA-Net. The results demonstrate near-infrared spectral imaging at 700-1000nm with 10nm resolution while achieving improvement of PSNR 1.47dB and SSIM 0.006.

## I. Introduction

Hyperspectral imaging acquires a series of spectral band images for a single scene, where each wavelength-specific image reveals the unique optical characteristics of its respective spatial location. These spectral characteristics reveal the chemical properties of different material compositions, enabling the discrimination of distinct substances. Leveraging these unique characteristics, hyperspectral imaging has found extensive applications across diverse fields, including remote sensing [1], [2], food industry [3], agriculture [4], [5], and cultural heritage [6].

The most straightforward methods of hyperspectral imaging employ dispersive elements (e.g., prisms) coupled with scanning mechanisms or utilize multiple narrow-band filters [7]. However, such systems are typically time-consuming and exhibit complex mechanical configurations, thus hindering their application in dynamic scene acquisition and lightweight deployment. Following the development of sensor technology and computational reconstruction algorithms, snapshot hyperspectral imaging systems, such as the coded-aperture snapshot spectral imager (CASSI) [8], [9], [10], have been proposed. The CASSI system achieves compressed spectral acquisition through coordinated encoding between dispersive optical elements and a mask, followed by computational reconstruction algorithms to recover the complete spectral datacubes. However, these methods still exhibit certain limitations. The system architecture remains inherently complex, posing challenges for integration. Furthermore, the use of a coding mask inevitably introduces reconstruction artifacts. Additionally, when operating in infrared spectral ranges, such systems require specially customized infrared optical components, contributing to cost disadvantages.

In recent years, spectral imaging systems based on novel materials, such as metasurface [11], [12], [13], photonic crystals [14], and Fabry-Pérot cavities [15], have garnered significant research attention. But these systems are constrained by both intricate optical design challenges and high manufacturing complexity, resulting in high cost and hindering mass applications. In contrast, diffractive imaging systems [16], [17], [18] provide a pathway to cost-effective integrated hyperspectral imaging based on point spread functions (PSFs) modulated designs and sensor-decoupled architecture.

In this work, we propose an integrated snapshot near-infrared hyperspectral imaging framework based on diffractive optics element (DOE). The proposed framework achieves a spectral resolution of 10 nm with 31 channels across 700 to 1000 nm in the near-infrared range. Furthermore, we introduce a neural network, termed NIRSA-Net. The NIRSA-Net demonstrates superior performance compared to state-of-the-art methods, achieving improvement of 1.47 dB in PSNR and 0.006 in SSIM. We anticipate that this work will provide practical insights for near-infrared hyperspectral imaging and holds significant potential for broad applications.

## II. Principle

### A. The Near-infrared Hyperspectral Imaging Framework

The integrated snapshot near-infrared hyperspectral imaging framework is shown in Fig. 1. Fig. 1(a) shows the diffractive encoder of the framework. Fig. 1(b) shows the generation process of the DOE. The DOE employs a rotationally symmetric distribution profile [12], parameterized by rotating a trainable 512×1 parameter. Fig. 1(c) shows the imaging model of the system. First, the point source light of $\lambda$ from $z$ could be modeled as:

$$U_1(x,y,\lambda,z) = \exp\left[i\frac{2\pi}{\lambda}\frac{x^2+y^2}{z}\right] \quad (1)$$

where $(x,y)$ represents the spatial coordinate. Second, the wave passes through the DOE, causing the changes of the amplitude and the phase. The wave is then changed to:

$$U_2(x,y,\lambda,z) = A(x,y)U_1 \exp\left[i\frac{2\pi}{\lambda}(n_\lambda - 1)h(x,y)\right] \quad (2)$$

where $A(x,y)$ is the amplitude function, which is typically treated as a constant $A_0$. $n_\lambda$ represents the refractive index of the DOE material for wavelength $\lambda$ and $h(x,y)$ represents the height map of the DOE. Next, the wave propagated to the sensor by the focal length $f$, resulting in the changes of the phase. The wave is changed to:

$$U_3(x,y,\lambda,z) = U_2 \exp\left[i\frac{2\pi}{\lambda}\frac{x^2+y^2}{2f}\right] \quad (3)$$

The PSFs of the system can be modeled as:

$$P_{psf}(x,y,\lambda,z) = |F\{U_3\}|^2 \quad (4)$$

We set the $z = z_0$ to simplify the calculation. The PSFs of the system is then convolved with the hyperspectral datacubes $I(x,y,\lambda)$ and mapped through the sensor response $R(\lambda)$ to generate the encoded image:

$$I_{encoded}(x,y) = \sum_{\lambda=\lambda_0}^{\lambda_k}(P_{psf}(x,y,\lambda,z_0) \otimes I(x,y,\lambda))R(\lambda) + n \quad (5)$$

where the $\lambda_0$ is the minimum wavelength of the hyperspectral datacubes and the $\lambda_k$ is the maximum wavelength. $n$ represents the sensor noise.

Fig. 1(d) shows the decoding process of the system. The encoded $I_{encoded}(x,y)$ is sent to the designed NIRSA-Net to obtain the reconstruction of the hyperspectral datacubes. Fig. 1(e) illustrates the structure of the system. The system consists of a DOE and a near-infrared camera, with the DOE fixed 50 mm in front of the camera sensor. The near-infrared camera is Baslar a2A2560-70umSWIR.

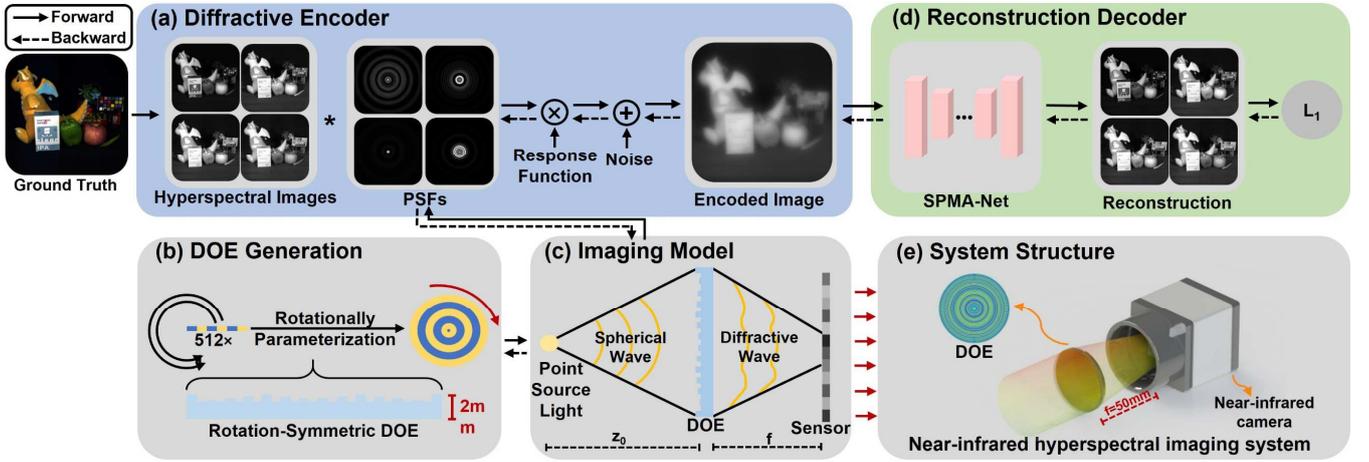

Fig. 1. The integrated snapshot near-infrared hyperspectral imaging framework. (a) Diffractive encoder. (b) DOE generation process. (c) Imaging model of the system. (d) Reconstruction decoder. (e) System structure.

*B. Near-infrared Spectral Imaging-Net*

The framework of the Near-infrared Spectral Attention-Net (NIRSA-Net) is shown in Fig. 2. Inspired by MST++ [20], we deployed a U-shaped Transformer-based architecture. The input of the NIRSA-Net is a H×W×1 encoded image from diffractive encoder. The encoded image is first processed through a 1×1 convolutional layer, producing feature maps $O_1$ of dimensions H×W×C, where C = 32. Then, $O_1$ passes through the embedding layer, which is denoted as a 3×3 convolutional layer. The output of the embedding layer is processed through $N_1$ Near-Infrared-Spectral Attention (NIR-SA) module, where $N_1$ = 2. The output with dimensions H×W×C is denoted as $O_2$. Next, $O_2$ is sent to a 4×4 strided convolutional layer as downsampling. The dimensions of feature maps are changed to (H/2)×(W/2)×(2C). Then, the feature maps pass through $N_2$ = 2 NIR-SA modules, generating the output feature maps $O_3$. $O_3$ undergoes downsampling module followed by processing through $N_3$ =2 PSA modules. The dimensions of feature maps are changed to (H/4)×(W/4)×(4C) and subsequently upsampled via a 2×2 strided transposed convolution layer, yielding the output feature maps $O_4$. $O_3$ and $O_4$ are jointly fed into a Near-infrared Fusion (NIR-Fusion) module to integrate deep and shallow features. The output feature maps are processed through $N_2$ NIR-SA modules, followed by a 2×2 strided transposed convolution, resulting in a upsampled feature maps $O_5$ of dimensions H×W×C. Then, $O_2$ and $O_5$ are jointly fed into a NIR-Fusion module. The output of NIR-Fusion is subsequently processed through $N_1$ NIR-SA modules, followed by a 3×3 convolutional layer. The output feature maps $O_6$ are then summed with $O_1$, then processed through a 1×1 convolutional layer followed by ReLU layer to generate the reconstruction.

The detailed architecture of NIR-SA and NIR-Fusion module are illustrated on the middle part of Fig. 2. Inspired by MST++, the NIR-SA module integrates Layer Normalization, Multi-head Self-Attention (MSA), Feed-Forward Network (FFN), and residual connections in a sequential architecture. The NIR-Fusion module employs multiple 3×3 convolutional layers to effectively integrate shallow and deep features through a three-stage process. Specifically, shallow features are first concatenated with deep features and processed through a 3×3 convolutional layer, combined with their own convolutional feature maps from a separate 3×3 convolution to generate an intermediate feature map. Second, deep features pass through a single 3×3 convolutional layer and are concatenated with the intermediate feature maps. Finally, this concatenated feature maps undergo a final 3×3 convolution to produce fusion output.

The detailed architecture of MSA and Feed Forward Network (FFN) module are illustrated on the bottom part of Fig. 2. The MSA module adopts the same multi-head attention computation as MST++, augmented with a 1×1 convolutional layer and 3×3 convolutional residual connections after the attention mechanism to enhance feature extraction. The FFN module maintains an identical structure to MST++.

The dataset of the network is acquired by commercial hyperspectral camera (Specim IQ). The loss function of NIRSA-Net is $L_1$ loss (Mean Absolute Error, MAE) between the reconstruction hyperspectral images and the ground truth. The network is trained for a total of 150 epochs. The learning rate of the network is 0.0005, which is multiplied by 0.9 every 30 epochs.

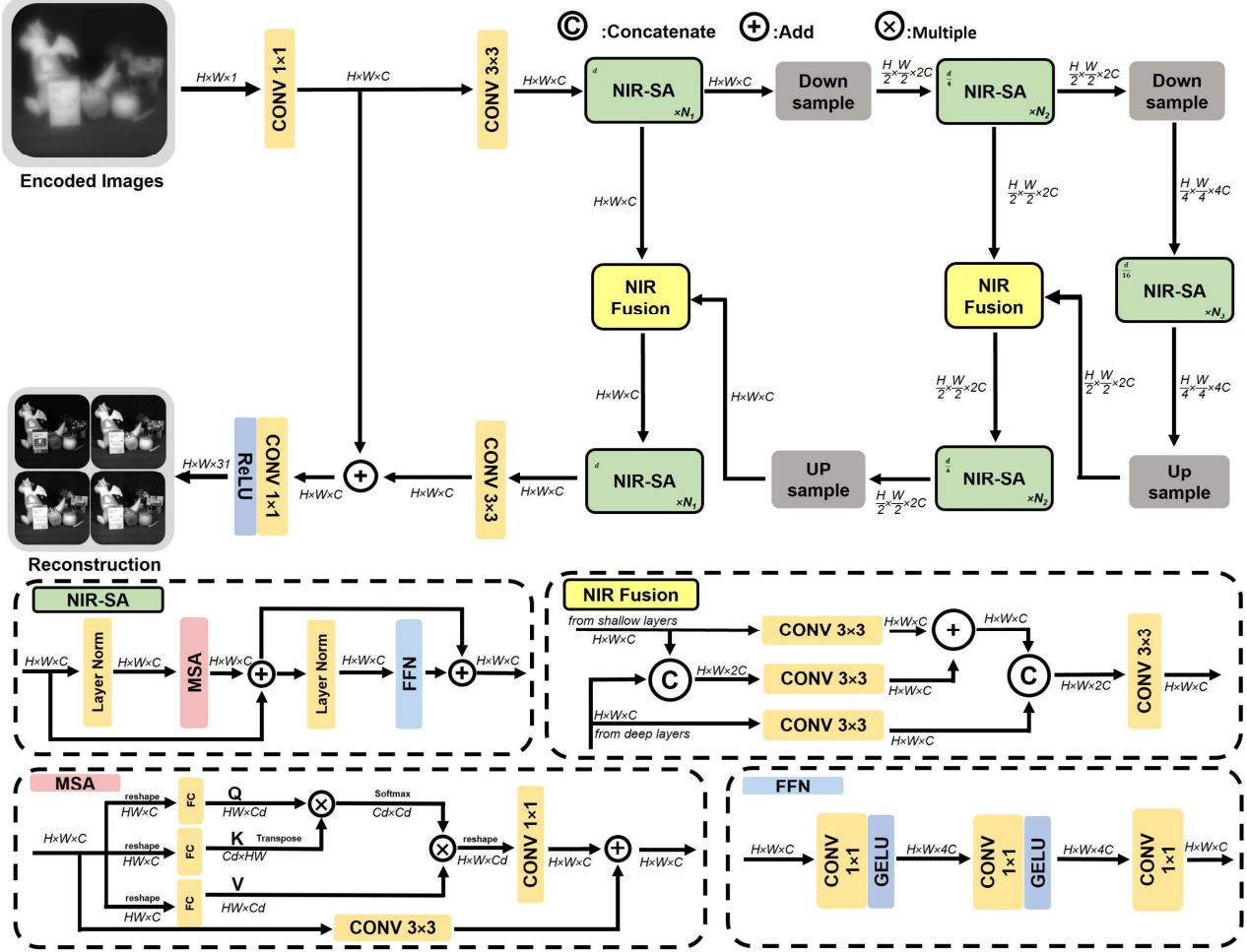

Fig. 2. The framework of NIRSA-Net.

## III. RESULTS

Fig. 3(a) shows the height map of the DOE. The DOE incorporates a 4 μm single-pixel pitch, with its 1024 × 1024 height map matrix quantized into 16 uniform steps, yielding a total structure depth of 2.2192 μm. The DOE has a 4.096 mm clear aperture with a 2 mm substrate thickness. The DOE incorporates a ±40 nm error per step to emulate the fabrication errors of lithography and reactive ion etching (RIE) techniques. Fig. 3(a) also shows the height maps of selected regions 1 and 2 for detailed visualization. The exhibited patterns demonstrate pronounced rotational symmetry, which aligns with our design. Fig. 3(b) shows the PSFs of the DOE across the wavelength range of 700-1000 nm.

In Fig. 4, we present a comparison of reconstruction performance between our proposed algorithm and existing state-of-the-art methods, including Res-Unet [19], HD-Net [21], MST++ [20]. The top part of Fig. 4 presents two selected scenes for spectral analysis.

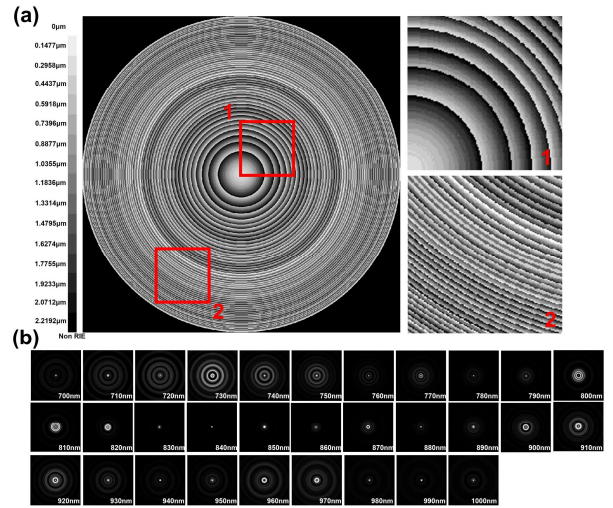

Fig. 3. The height map and the PSFs of the DOE. (a) The height map of DOE. (b) The PSFs of the DOE across the wavelength range of 700-1000 nm.

We sampled two representative points (Point 1 and 2) and analyzed their spectral signatures. Both points exhibit red coloration in the visible spectrum, corresponding to a tomato surface and a red 'Thorlabs' box respectively. Their spectral signatures diverge significantly in the near-infrared region. The spectral curves of Point 1 and 2 are shown on the top part of Fig. 4. The spectral curves of Point 1 and 2 demonstrate that our algorithm achieves superior accuracy compared to alternative methods, precisely capturing both the characteristic reflectance decline of tomato surfaces over 900 nm and maintaining the smooth spectral trend of the 'Thorlabs' box in the near-infrared region. The bottom part of Fig. 4 presents reconstructed images of multi bands reconstructed from different methods. We selected a representative region (orange boxed region) in 710nm image of scene 1 for spatial detail comparison. Compared to the images reconstructed by other methods, which exhibit reconstruction failures and luminance discontinuities, our method performs results with superior smoothness that show strong agreement with the ground truth. The results demonstrate that our algorithm maintains better performance in spatial reconstruction fidelity. Our method achieves a PSNR of 35.70 dB and SSIM of 0.977 on the test dataset, representing improvements of 1.47 dB and 0.006, respectively, over state-of-the-art algorithms.

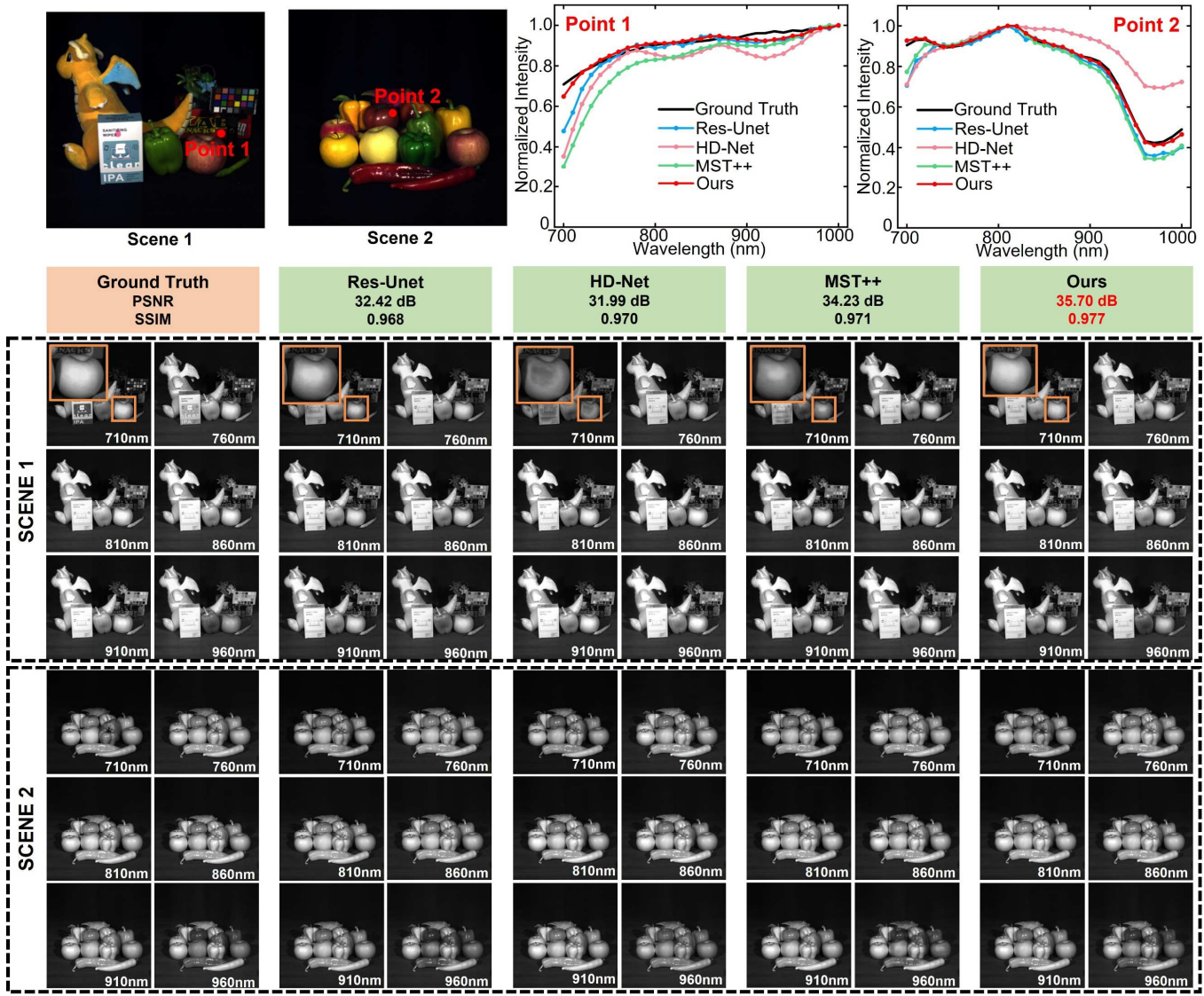

Fig. 4. The comparison of reconstruction results between NIRSA-Net and existing state-of-the-art methods.

## IV. CONCLUSION

In this work, we propose an integrated snapshot near-infrared hyperspectral imaging framework. The framework consists of a designed DOE and a novel algorithm called NIRSA-Net, achieving hyperspectral imaging with 10nm resolution across 31 spectral bands in the 700-1000nm near-infrared range. Comparative studies with state-of-the-art algorithms demonstrate our system's superior fidelity in both spectral and spatial reconstruction, specifically showing improvements of 1.47dB in PSNR and 0.006 in SSIM. These results confirm the strong potential of this work for wide applications such as remote sensing, agricultural monitoring, and material characterization.

ACKNOWLEDGMENT

This work was supported by the Science and Technology Innovation Project for Xiong'an New Area (NO. 2023XAGG0089); National Natural Science Foundation of China (No. 62522502, 62371056); Major Science and Technology Support Program of Hebei Province (No.